\documentclass[useAMS,usenatbib]{mn2e}

\usepackage{times,graphicx,amssymb,natbib, amsmath}
\pdfoutput=1

\newcommand{\msol}{M_{\rm \odot}}

\title[Classifying structures in SPH]{Tensor classification of structure in smoothed particle hydrodynamics density fields}
\author[Duncan Forgan, Ian Bonnell, William Lucas and Ken Rice]{Duncan Forgan $^{1}$\thanks{E-mail:dhf3@st-andrews.ac.uk}, Ian Bonnell$^{1}$, William Lucas$^{1}$ and Ken Rice$^{2}$  \\
$^{1}$Scottish Universities Physics Alliance (SUPA), School of Physics and Astronomy, University of St Andrews, North Haugh, St Andrews KY16 9SS \\
$^{2}$SUPA, Institute for Astronomy, University of Edinburgh, Blackford Hill, Edinburgh EH9 3HJ \\
}

\begin{document}

\date{Accepted}

\pagerange{\pageref{firstpage}--\pageref{lastpage}} \pubyear{}

\maketitle

\label{firstpage}

\begin{abstract}

\noindent As hydrodynamic simulations increase in scale and resolution, identifying structures with non-trivial geometries or  regions of general interest becomes increasingly challenging.  There is a growing need for algorithms that identify a variety of different features in a simulation without requiring a ``by-eye'' search.  We present tensor classification as such a technique for smoothed particle hydrodynamics (SPH).  These methods have already been used to great effect in $N$-Body cosmological simulations, which require smoothing defined as an input free parameter.  We show that tensor classification successfully identifies a wide range of structures in SPH density fields using its native smoothing, removing a free parameter from the analysis and preventing the need for tesselation of the density field, as required by some classification algorithms.  As examples, we show that tensor classification using the tidal tensor and the velocity shear tensor successfully identifies filaments, shells and sheet structures in giant molecular cloud simulations, as well as spiral arms in discs.  The relationship between structures identified using different tensors illustrates how different forces compete and co-operate to produce the observed density field.  We therefore advocate the use of multiple tensors to classify structure in SPH simulations, to shed light on the interplay of multiple physical processes.

\end{abstract}


\section{Introduction}


\noindent Numerical simulations allow astrophysicists to study and catalogue physical processes on spatial and temporal scales inaccessible to observations.  Flows of matter, energy and angular momentum produce a variety of structures, some of which are simple to detect ``by eye'', for example spiral arms in astrophysical discs (e.g. \citealt{Forgan2011, Dipierro2015}), filaments of turbulent gas in giant molecular clouds \citep{Klessen2000,Krumholz2011,Bonnell2013} or the network of dark matter that constitutes the Cosmic Web (e.g. \citealt{Hahn2007}).

However, ``by eye'' approaches are unsuitable for determining the status of fluid elements on the periphery of a structure - does such a simulation element ``belong'' to a filament or not? Also, incorporating humans directly into any computational algorithm is time-inefficient.  For very large simulations, it is increasingly crucial that any structures produced can be described quickly and reliably.  Structure identification is important not only for comparing with observational data, but also to understand the mechanisms that drive structure formation, such as self-gravity, turbulence and shocks.

Particle-based simulations have been particularly successful in computational astrophysics, and structure identification has long been a consideration for simulators in this area.  $N$-Body calculations of structure growth in the early Universe have led to useful topological classifications of the dark matter distribution \citep{Hahn2007, Forero-Romero2009, Sousbie2011,Hoffman2012, Libeskind2013a} as well as identification of dark matter halos using halo-finding algorithms such as the friends-of-friends (FoF) algorithm, which has many versions and guises (\citealt{Davis1985}, \citealt{knollmann-knebe2009} and references therein).  

In other astrophysical regimes, such as star and planet formation, the local dark matter distribution is less relevant.  In these circumstances baryonic physics, especially hydrodynamics, dominates, but particle-based simulation is still a commonly used tool, and the need for structure identification remains.  Smoothed Particle Hydrodynamics (SPH) is a Lagrangian particle simulation technique, which approximates a fluid by a distribution of particles.  Detailed reviews of its governing algorithms can be found in \citet{Monaghan_92,Monaghan_05} and \citet{Price2012}, but for completeness we will state its fundamentals here.   Each particle is assigned a mass $m_i$, position $\mathbf{r}_i$, velocity $\mathbf{v}_i$ and internal energy $u_i$, and a smoothing length, $h_i$.  From this, the properties of the fluid at any location are derived via interpolation of the particle's properties, applying a smoothing kernel $W$ to each contribution.  For example, the local volume density at a location $\mathbf{r}$ can be determined via

\begin{equation}
\rho (\mathbf{r}) = \sum^{N}_{i=1} m_i W(\left|\mathbf{r} - \mathbf{r}_i\right|, h),
\end{equation}

\noindent where $N$ represents the number of particles contributing to the sum.  Smoothing kernels typically have compact support, and hence are zero at separations greater than $2h$.  $N$ reduces to the number of ``nearest neighbours'' that reside within $2h$ of the particle.  This density estimate - along with a suitable definition for the system Lagrangian/Hamiltonian and application of the Euler-Lagrange/Hamilton's equations - is sufficient to derive equations of motion for the entire system, and (with appropriate measures taken for treatment of shocks and mixing) can simulate fluids in any geometric configuration.  Indeed, it has been this ability to deal with a wide variety of geometries that has made SPH such a popular tool in astrophysics, as well as in many other fields of research.  

But how should SPH users attempt to classify structure in their simulations? Some approaches have the benefit of simplicity, for example identifying contiguous regions using a density cut has been effective in use cases such as finding giant molecular clouds in SPH simulations of galactic discs (e.g. \citealt{Dobbs2014}).  An obvious counter-argument to such approaches is what justifies the value of the density cut used.  Depending on the system geometry, this value can be motivated by observational constraints, or even analytically derived, but in general this may not be the case.

Several algorithms use the linked lists of SPH neighbours, in a vein quite similar to FoF algorithms.  The CLUMPFIND algorithm, originally developed for imaging purposes \citep{Williams1994}, identifies separate objects in a simulation according to global maxima in density, or global minima in the gravitational potential \citep{Klessen2000}.  Given an appropriate prescription for dealing with objects that touch or overlap in terms of particle population, clumpfind techniques have been useful for identifying fragments both in giant molecular clouds \citep{Smith2009} and also in discs (Hall et al, in prep).  Also of note is the use of minimum spanning trees in hybrid SPH/$N$-Body calculations \citep{Maschberger2010,Kruijssen2012}.  However, this technique generally detects structure in the $N$-Body particles rather than the gas.  Finally, we must mention the DisPerSE algorithm,  which uses Morse theory to classify structure using the Delaunay triangulation of discrete data \citep{Sousbie2011}.  This has proven to be extremely effective in classifying structures both in numerical simulations (e.g. \citealt{Smith2014}) and in observational data (e.g. \citealt{Arzoumanian2011}) but its use in SPH would require tesselation of the density field, an extra layer of assumptions which is preferentially avoided.

In this paper, we demonstrate the simplicity and power of using \emph{tensor classification} methods on pure SPH data.  Originally conceived and applied to pure $N$-Body simulations of the cosmic Web\footnote{This technique has both numerical and observational applications, cf \citet{Guo2015}'s study of the distribution of galaxies}, we show that these methods can be applied to pure SPH simulations, with the advantage of fewer free parameters.  Judicious use of tensor eigenvalues and eigenvectors can not only identify which particles belong to a given structure, but also the orientation of that structure with respect to its environment.  Classifying via multiple tensors allows finer structures to be elucidated, and gives insight into how various physical forces interact to produce the underlying structures seen. 

We structure this paper as follows: in section \ref{sec:methods} we describe the two tensors we use to classify SPH structures; in section \ref{sec:tests} we test their ability to classify analytically soluble structures; in section \ref{sec:applications} we display some of the many applications of these techniques, and in sections \ref{sec:discussion} and \ref{sec:conclusions} we discuss and summarise the work.

\section{Methods }\label{sec:methods}

\subsection{The Tidal Tensor}


The tidal tensor $T_{ij}$ is simply the Hessian of the gravitational potential, $\phi$.  In Cartesian co-ordinates $\{x_i\}$:

\begin{equation}
T_{ij} = \frac{\partial^2 \phi }{ \partial x_i \partial x_j}.
\end{equation}

\noindent Note that $\phi$ is the standard gravitational potential, which solves the canonical Poisson equation:

\begin{equation}
\nabla^2 \phi = 4\pi G \rho,
\end{equation}

\noindent rather than the pseudo-potential commonly used in cosmological simulations to solve for the matter overdensity. Strictly, $T_{ij}$ as defined above is the \emph{deformation tensor}, as the tidal tensor refers to the traceless component of the Hessian.  These labels are commonly interchanged in the context of structure identification \citep{Hahn2007,Forero-Romero2009}.  We refer to it as the tidal tensor to be clear as to its differences from the velocity shear tensor, which we describe in the next section.  We utilise the dimensionless form of this tensor:

\begin{equation}
T_{ij} = \frac{\partial^2 \phi }{ \partial x_i \partial x_j} \frac{h^2}{\left|\phi\right|}.
\end{equation}

\noindent We elect a normalisation based on smoothing length as derivatives are computed on the smoothing kernel, and are hence resolution-dependent.  In its simplest form:

\begin{equation}
\nabla F (\mathbf{r}) = \sum^{N}_{i=1} \frac{F_i m_i}{\rho_i} \nabla W(\left|\mathbf{r} - \mathbf{r}_i\right|, h).
\end{equation}

\noindent As a result, this places resolution limits on the gradients resolvable by SPH simulations.  In the weak field limit, we expect derivatives to follow

\begin{equation}
\frac{\partial F}{\partial x_i \partial x_j} \sim \frac{F}{h^2},
\end{equation}

\noindent and hence corresponding tensor eigenvalues will be of that order.  Our normalisation is therefore a depiction of the strength of the measured gradients in potential \emph{at the local simulation resolution}, which will become useful later.

\subsection{The Velocity Shear Tensor}

The velocity shear tensor, $\Sigma_{ij}$, is:

\begin{equation}
\Sigma_{ij} = -\frac{1}{2}\left(\frac{\partial v_i}{\partial x_j} + \frac{\partial v_j}{\partial x_i} \right).
\end{equation}

\noindent The astute will note this is equivalent to the strain rate tensor, which measures how the mean velocity in the medium changes between two (infinitesimally) separate locations. The negative sign indicates that we are interested in compression of the medium.  When used in cosmological comoving co-ordinates, it is customary to make this tensor dimensionless by adding in a factor of $1/H_0$, where $H_0$ is the Hubble constant \citep{Hoffman2012}.  As with the tidal tensor above, we make $\Sigma$ dimensionless using the local smoothing length:

\begin{equation}
\Sigma_{ij} = -\frac{1}{2}\left(\frac{\partial v_i}{\partial x_j} + \frac{\partial v_j}{\partial x_i} \right) \frac{h}{|\mathbf{v}|}.
\end{equation}

\noindent Again, we adopt an $h$-dependent normalisation due to the $h$-dependence of the derivative.

\subsection{Tensor Classification}

\noindent In all cases, the classification of tensors is algorithmically identical.  Firstly, the tensor's eigenvalues $\lambda_i$ and their corresponding eigenvectors $\mathbf{n_i}$ are computed:

\begin{equation}
T \mathbf{n}_j = \lambda_j \mathbf{n}_j
\end{equation}

\noindent We label the eigenvalues such that $\lambda_1 \geq \lambda_2 \geq \lambda_3$.  The tensors used in this work are real and symmetric, and hence their eigenvalues are always real.  Both tensors assume that the fields being investigated (potential, velocity) are smooth and continuous, so that the derivatives are always defined.  In $N$-Body calculations, this means the fields must first be smoothed, with the smoothing scale $R_s$ a free parameter.  SPH simulations enjoy the advantage of being already adaptively smoothed according to the local smoothing length $h$.  Our calculations henceforth do not require any extra smoothing, and by extension they do not need the above free parameter.

In the case of the tidal tensor, we can now appeal to Zeldovich theory \citep{Zeldovich1970} to motivate our classification.  By considering a test particle in orbit around a local extremum in the potential ($\nabla \phi=0$), we can linearise its equation of motion to:

\begin{equation}
\ddot{x}_i = -T_{ij} x_j.
\end{equation}

\noindent We neglect pressure forces and other terms in this equation of motion for the purposes of this illustration.  As we can select an appropriate basis such that $T_{ij}$ is diagonal, the linear motion of the particle is governed by the sign of the eigenvalues.  If none of the eigenvalues are positive, then the orbit cannot be stable under any configuration.  If all the eigenvalues are positive, the orbit is stable under all configurations.  If one or two eigenvalues are positive, then some configurations are stable and some are not.  More rigorously, the number of positive eigenvalues $E$ describes the dimension of the local stable manifold.  We should therefore expect tidal tensor classifications to be effective in regions where pressure gradients are weak compared to potential gradients.

In a cosmological context, the above argument can also be made for the velocity shear tensor, as the velocity field and gravitational fields are close to identical in this limit in the linear regime, up to a scaling dependent on cosmological parameters  \citep{Hoffman2012}.  We are not always at the liberty of making this argument, but we can state that calculating $E$ for $\Sigma_{ij}$ allows us to diagnose the dimension of the local flow at that instant, albeit with no indication as to whether the configuration is dynamically stable or gravitationally bound.

If $E$ can be calculated for our tensor of choice, we can then make the following classification:

\begin{itemize}
\item $E=0 \rightarrow $ ``void'' (0-D manifold)
\item $E=1 \rightarrow $ ``sheet'' (1-D manifold)
\item $E=2 \rightarrow $ ``filament'' (2-D manifold)
\item $E=3 \rightarrow $ ``cluster'' (3-D manifold)
\end{itemize}

\noindent In simple terms, $E$ is equivalent to the number of dimensions the local medium is being ``squeezed'' in.  Sheets are produced by squeezing along one axis, filaments along two axes, clusters along three.  Classification via $T$ is dynamical, and indicates the manifold dimension of a collapsing region.  Classification via $\Sigma$ is kinematic, and indicates the manifold dimension of a flow that can either be gravitationally bound or unbound.

We have further information to glean from the tensor, in the form of the corresponding eigenvectors.  The single positive eigenvalue of a sheet ($\lambda_1$) has a corresponding eigenvector ($\mathbf{n}_1$) which is parallel to the normal of the sheet.  The single negative eigenvalue of a filament ($\lambda_3$) has a corresponding eigenvector ($\mathbf{n}_3$) which is parallel to the flow direction of the filament.  This data now allows us to compare alignments of sheets and filaments to the local environment.

However, we should be cognisant of the dangers of attempting to discriminate between structure using small, non-zero values of quantities in numerical simulations.  \citet{Forero-Romero2009} give an example of attempting to classify void particles at the interface between a void and a sheet region (for example).  According to the above algorithm, a particle with a single (but infinitesimally small) positive eigenvalue will be classified as a sheet.  This is equivalent to suggesting that in the region of that particle, the local medium will collapse into a 1D manifold with a symmetry axis given by the particle's $\mathbf{n}_1$ eigenvector.  In practice, the collapse timescale implied by such a small eigenvalue is so long that it is unlikely to occur in the simulation, given the other local timescales at play.  \citet{Forero-Romero2009} claim that this explains the relatively low proportion of voids found by \citet{Hahn2007} when using the positive/negative eigenvalue criterion detailed above.

A solution to this is to instead define $E$ as the number of eigenvalues that exceed some threshold value $\lambda_T$.  Use of $\lambda_T$ allows us to ameliorate not only this issue of floating point accuracy, but general issues that might arise from using this algorithm on SPH density fields, such as particle disorder and poor resolution in regions of low particle number.  

This threshold parameter $\lambda_T$ is the only free parameter in our classification scheme.  In some simple cases, the value of the threshold parameter can be estimated from analytic considerations of the above collapse timescale (see Appendix A of \citealt{Forero-Romero2009} and also \citealt{Alonso2015}).  In the case of this work, we investigate the behaviour of the classification as $\lambda_T$ is varied.  Thanks to our use of the local smoothing scale in the normalisation of both tensors, $\lambda_T$ is now a local threshold, and as a result we find a common property of structure classification as a function of $\lambda_T$.  As we shall see throughout, provided that $\lambda_T <<1$, structures are well resolved.  This puts SPH at a distinct advantage to pure $N$-Body simulations, which often rely on multiscale smoothing approaches to ensure appropriate structure resolution using a global threshold (e.g. \citealt{Hoffman2012}).

\section{Tests}\label{sec:tests}

\noindent We now consider some simple test cases for both tensors.  These test cases have analytic solutions, and therefore we can compare them directly to simulation data.  Throughout this paper, SPH simulations are run using the \texttt{sphNG} code \citep{Bate_code}, using a variety of prescriptions for the gas equation of state and radiative transfer, which we will discuss where appropriate.

\subsection{The Tidal Tensor of a Uniform Density Sphere}

\noindent Consider a sphere, with radius $R$ and uniform density $\rho_0$, embedded in a vacuum ($\rho(r) = 0$ for $r>R$) . 

\subsubsection{Analytic Solution}

\noindent The potential for this system is spherically symmetric, and hence:

\begin{equation}
\phi(\mathbf{r}) =\phi(r) = -4\pi G \left( \int^{r}_{0} \rho(r') r'^2 dr' + \int^{\infty}_{r} \rho(r') r' dr'\right).
\end{equation}

\noindent This integral has the solution:

\begin{equation}
\phi(r) = \begin{cases}
\frac{-GM}{2R^2} (3R^2 - r^2) & r \leq R \\
 \frac{-GM}{r}& r >R
\end{cases}
\end{equation}

\noindent Where we have substituted for the total mass of the sphere: $M = 4/3 \pi G \rho_0 R^3$.  As the problem is spherically symmetric, we can use the greatly simplified expression for $T_{ij}$ in spherical polar coordinates:

\begin{equation} 
T_{ij} = diag\left[\frac{\partial^2 \phi}{\partial r^2}, \frac{1}{r}\frac{\partial \phi}{\partial r}, \frac{1}{r}\frac{\partial \phi}{\partial r}\right].
\end{equation}

\noindent In the case $r \leq R$:

\begin{equation}
T_{ij} = \frac{GM}{R^3} \delta_{ij}.
\end{equation}

\noindent All the eigenvalues are equal, and positive.  Therefore at any location inside $R$, the tidal tensor possesses an $E=3$, and that location will hence be classified as a cluster.  This is relatively intuitive - particles inside the sphere experience gravitational forces in all three dimensions. As $r$ is increased beyond $R$, the gravitational force becomes preferentially directed back towards the sphere, and the particles are ``aware'' of the two dimensional nature of the sphere's surface.  The tidal tensor takes the form:

\begin{equation}
T_{ij} = \frac{GM}{r^3} diag[-2, 1, 1].
\end{equation}

\noindent In this case, $E=2$ and this region is classified as a filament.  As $r\rightarrow \infty$, the force tends to zero, as does eigenvalues of the tensor.  As a result, $E\rightarrow0$ and the classification tends towards void status.

\subsubsection{SPH Results}

\noindent On the one hand, this test is equivalent to testing the ability of SPH particles to resolve a surface.  On the other hand, this tests the ability of SPH particles to simulate a uniform density object.

We conduct this test using two sets of SPH initial conditions.  In the first, 500,000 particles are placed randomly in an unrelaxed distribution, resulting in relatively significant particle disorder; in the second, 300,000 particles are distributed on a lattice with equal separations.  These two cases represent to some degree the best and worst case scenarios in SPH simulations.  These snapshots are not evolved - we are interested merely in the classification of the SPH particles at this given instant.  Over the course of a simulation, SPH particles will arrange themselves to minimise disorder in the system \citep{Monaghan_05,Price2012}, resulting in conditions belonging somewhere in between the cases we study here\footnote{Note that this ``intrinsic re-meshing'' will only occur for specific formulations of SPH - in effect, whether one derives the SPH equations from the Lagrangian or from the Hamiltonian.  As a result, most SPH codes select the latter, as it preserves intrinsic re-meshing. \citet{Price2012}'s discussion of this is excellent, and hence we will not repeat it here}.

\begin{figure*}
\begin{center}$\begin{array}{cc}
\includegraphics[scale=0.4]{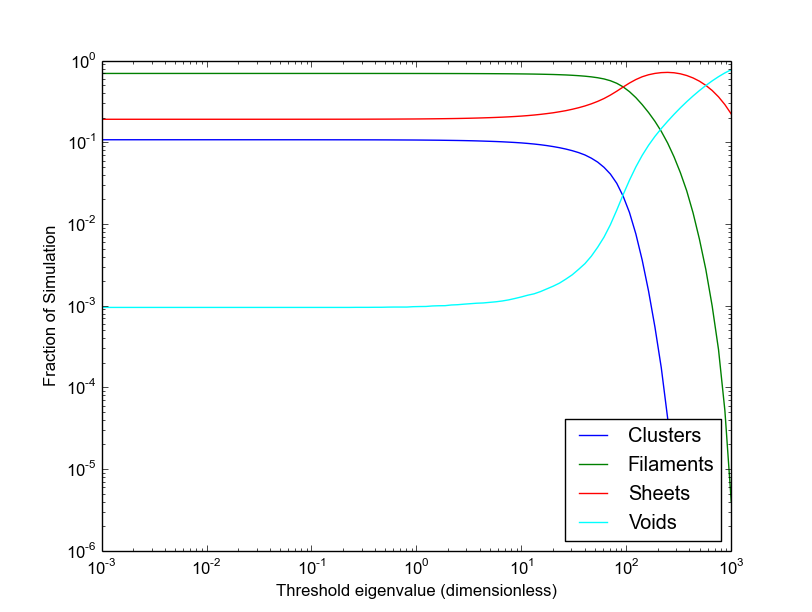} &
\includegraphics[scale=0.4]{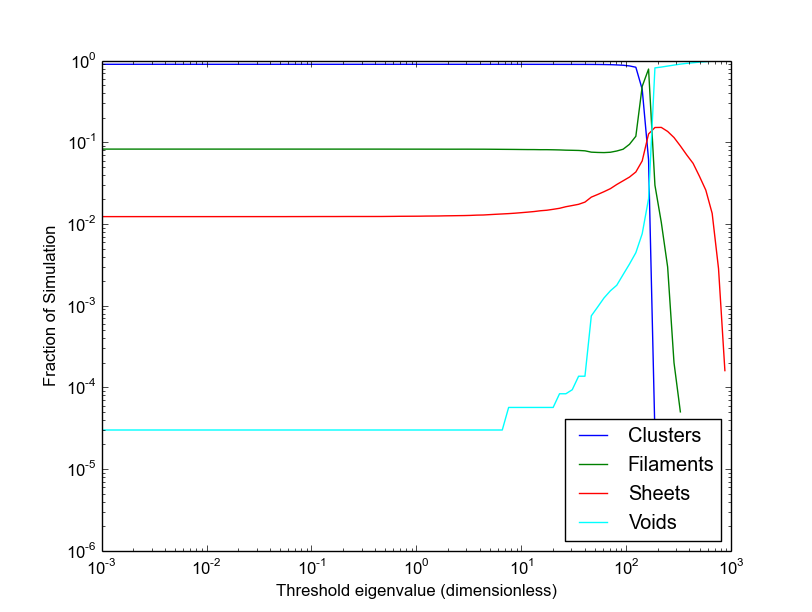} \\
\end{array}$
\caption{The uniform sphere test of the tidal tensor.  The fraction of total particles in each class is plotted as a function of the threshold parameter, $\lambda_T$ for (left) Case 1, an unrelaxed, noisy particle distribution, and (right) Case 2, a lattice-based particle distribution.\label{fig:tidaltest_vsthreshold}}
\end{center}
\end{figure*}

\begin{figure*}
\begin{center}$\begin{array}{cc}
\includegraphics[scale=0.4]{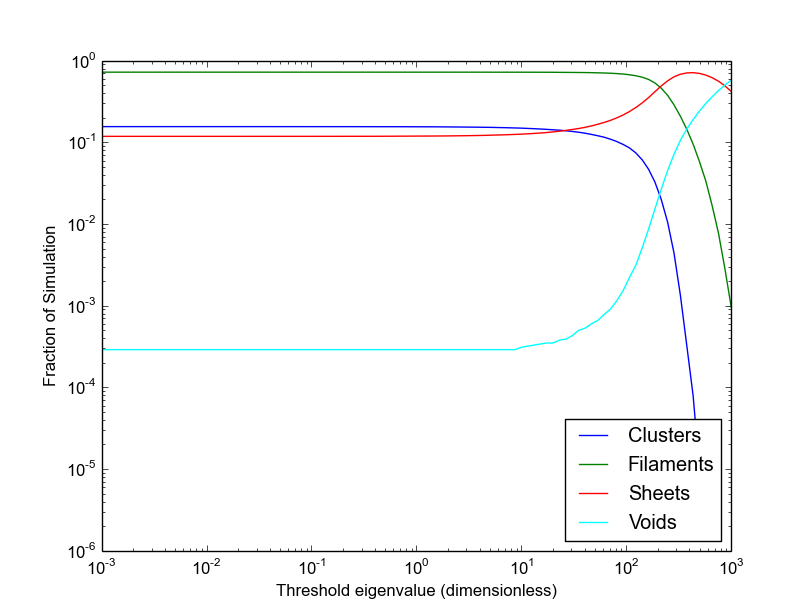} &
\includegraphics[scale=0.4]{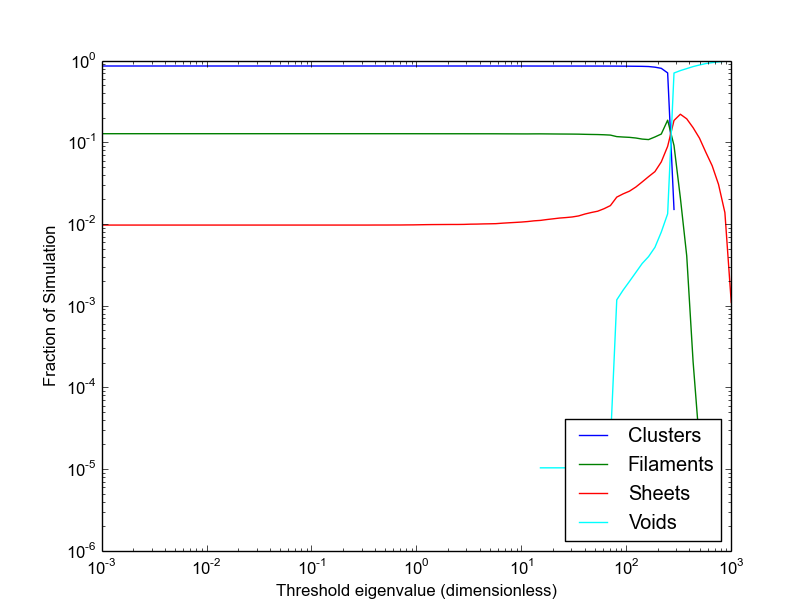} \\
\end{array}$
\caption{As Figure \ref{fig:tidaltest_vsthreshold}, but with a lower particle number (100,000).  This demonstrates that the appropriate threshold to adopt is $\lambda_T<<1$, and that this is resolution independent.  \label{fig:tidaltest_1e5}}
\end{center}
\end{figure*}

A successful test will identify almost all SPH particles as clusters, with a small handful identified as filaments depending on the simulation's ability to resolve the spherical surface.  Figure \ref{fig:tidaltest_vsthreshold} shows the resulting classifications for both snapshots as a function of the threshold value, $\lambda_T$.  In both cases, we can see that $\lambda_T$ must be kept sufficiently small for the classification to be functional: once $\lambda_T$ is larger than the mean eigenvalue, all particles are classed as voids.  Below a critical value of $\lambda_T \sim 1$, the classification system is attempting to probe the sub-$h$ limit.  In the case of the tidal tensor, the potential in the sub-$h$ limit is kernel-smoothed, and hence the fractions of particles in each class converges to a fixed value.  The scaling of the tensor (and subsequently its eigenvalues) by the local smoothing length ensures that the critical value of $\lambda_T$ is independent of resolution.  We confirm this by running both cases again at lower resolution (100,000 particles), with similar results (Figure \ref{fig:tidaltest_1e5}).  

\begin{figure*}
\begin{center}$\begin{array}{cc}
\includegraphics[scale=0.3]{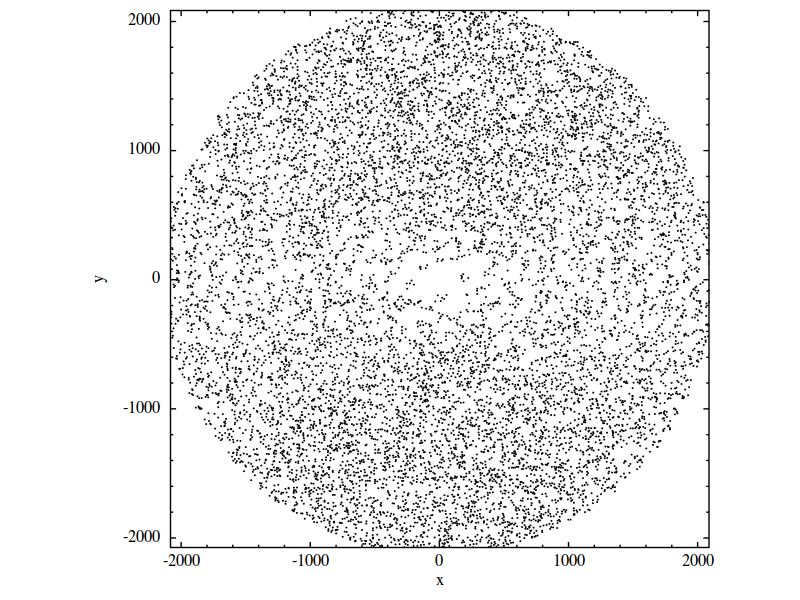} &
\includegraphics[scale=0.3]{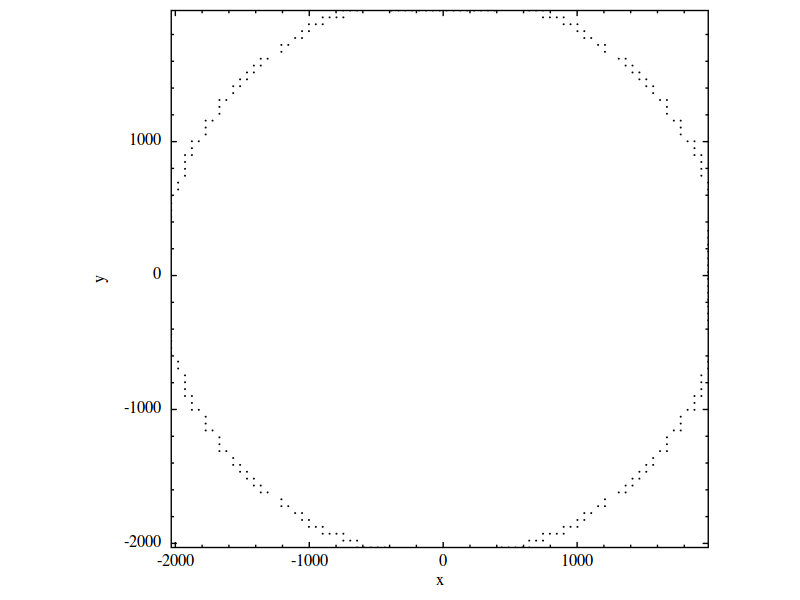} \\
\end{array}$
\caption{The distribution of particles classified as filaments in the higher resolution simulations of Case 1 (left) and case 2 (right), for a threshold parameter of $\lambda_T=10^{-6}$.  The above figures show a cross section of each simulation, centred on $z=0$ with a thickness of 100.\label{fig:tidaltest_plots}}
\end{center}
\end{figure*}

The unrelaxed snapshot (Case 1) shows the danger of particle disorder for this classification system.  The sphere's inhomogeneities ensure that many particles are classified as filaments or sheets rather than clusters.  This is not merely a surface feature problem, as filament and sheet particles reside at all distances from the centre of the sphere (left panel of Figure \ref{fig:tidaltest_plots}).

In the lattice snapshot (Case 2), the lack of disorder in the system allows the tidal tensor to correctly identify over 90\% of the particles as cluster objects.  The relatively low resolution of the sphere's surface (with thickness $\sim h$) results in filament identifications for some 8\% of particles (right panel Figure \ref{fig:tidaltest_plots}). 

\noindent What should we conclude from this test? A well-ordered particle distribution can pass this test with ease, but noisy distributions cannot.  We should therefore consider our particle classifications carefully, and apply this technique to systems where we can be confident that any residual noise from initial conditions has been well dissipated in the subsequent evolution of the simulation.

\subsection{The Velocity Shear Tensor for a radially expanding shell \label{sec:velsheartest}}

\noindent Consider a radially expanding shell of gas at uniform density with velocity $v_1$, into a medium with velocity $v_0 << v_1$. The velocity gradient is entirely determined by the radial velocity,

\begin{equation}
\nabla \mathbf{v} = \frac{\partial v_r}{\partial r} \mathbf{\hat{r}},
\end{equation}

\noindent and all other velocity components are zero.

\subsubsection{Analytic Solution}

\noindent To obtain this solution, we should rewrite the velocity shear tensor in spherical polar co-ordinates.  This gives a quite  unwieldy general expression, but mercifully when we apply our constraints: 

\begin{equation}
v_{\theta} = v_{\phi} = \frac{\partial v}{\partial \theta} = \frac{\partial v}{\partial \phi}=0,
\end{equation}

\noindent the tensor becomes diagonal:

\begin{equation}
\Sigma_{ij} = diag\left[-\frac{\partial v_r}{\partial r}, -\frac{v_r}{r}, -\frac{v_r}{r}\right].
\end{equation}

\noindent As the radial velocity is positive, and the velocity gradient is negative, this gives $E=1$, and therefore the expanding shell is classified as a sheet.  Outside of the expanding shell, the velocity gradient is zero, and hence $E=0$, resulting in a void classification.

\subsubsection{SPH Results}

\noindent To test this solution, we begin with a cube of 500,000 particles, of side $L=1000$.  Particles with a radius $R$ from the centre of the cube are assigned a radial velocity $v_1$, and particles outside of this radius are assigned a radial velocity $v_0=0.8v_1$.  We select a non-zero value for $v_0$ to demonstrate the tensor classification algorithm is truly Lagrangian.

Note that our setup is not an expanding shell, but an expanding sphere.  In the above analytic solution, we should only expect particles that sense a non-zero velocity gradient locally, i.e. those particles on the boundary between the sphere and the surrounding medium, to be classified as sheet particles.  Inside and outside the sphere, we should expect zero velocity gradients, and hence we should obtain void classifications for those particles.

Again, we run two tests of this case.  In the first, we generate a noisy cube of particles by placing them randomly in Cartesian co-ordinates, and in the second we generate a cubic lattice.  Particles inside a radius of $R=500$ from the centre are assigned a radial velocity $v_r = v_1 = 10$, and all other particles are assigned a radial velocity $v_0 = 8$.

\begin{figure*}
\begin{center}$\begin{array}{cc}
\includegraphics[scale=0.4]{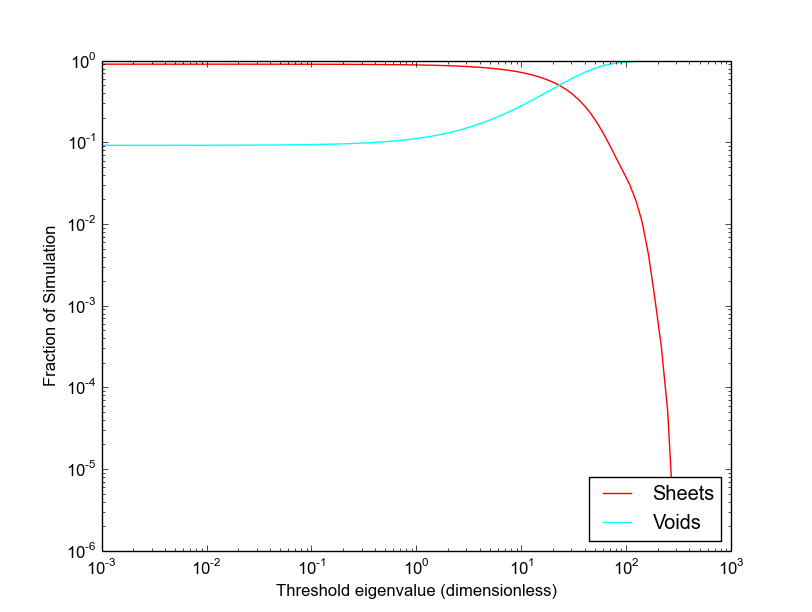} &
\includegraphics[scale=0.4]{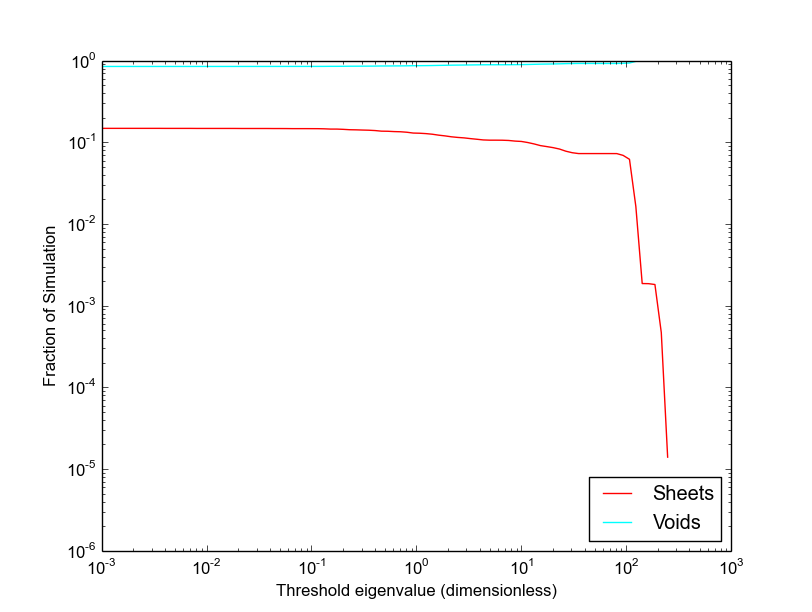} \\
\end{array}$
\caption{The expanding sphere test of the velocity shear tensor.  The fraction of total particles in each class is plotted as a function of the threshold parameter, $\lambda_T$ for (left) Case 1, an unrelaxed, noisy particle distribution, and (right) Case 2, a lattice-based particle distribution.\label{fig:veltest_vsthreshold}}
\end{center}
\end{figure*}

\begin{figure*}
\begin{center}$\begin{array}{cc}
\includegraphics[scale=0.4]{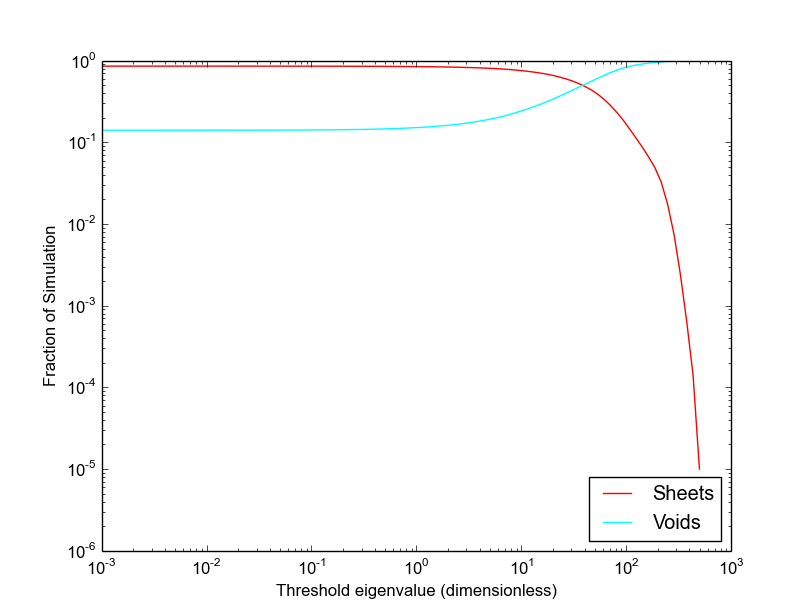} &
\includegraphics[scale=0.4]{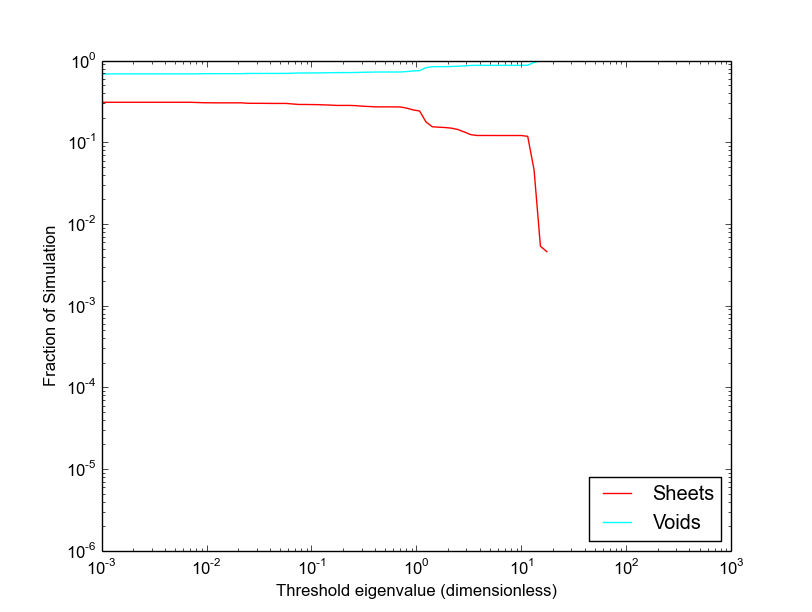} \\
\end{array}$
\caption{As Figure \ref{fig:veltest_vsthreshold}, but with a lower particle number (100,000).  Again, this demonstrates that the appropriate threshold to adopt is $\lambda_T<<1$, and that this is resolution independent.  \label{fig:velltest_1e5}}
\end{center}
\end{figure*}

Figure \ref{fig:veltest_vsthreshold} shows how the global classification of particles varies for both cases as the eigenvalue threshold $\lambda_T$ is increased.  In both cases, at appropriate values of $\lambda_T<1$, the majority of particles are identified as sheets and voids.  In the random case (left panel of Figure \ref{fig:veltest_vsthreshold}), particle noise results in an excess of sheet classifications, even at low $\lambda_T$.  In the lattice case, there is no particle noise, and hence there are no erroneous classifications of clusters and filaments at any $\lambda_T$.  As before, we demonstrate that the $\lambda_T$ condition is resolution independent by repeating the test at lower resolution (Figure \ref{fig:velltest_1e5}).

\begin{figure*}
\begin{center}$\begin{array}{cc}
\includegraphics[scale=0.3]{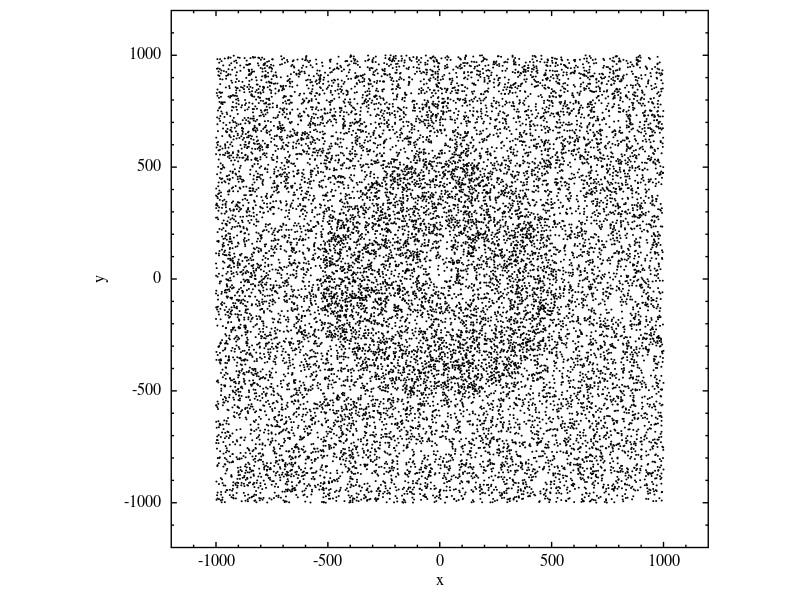} &
\includegraphics[scale=0.3]{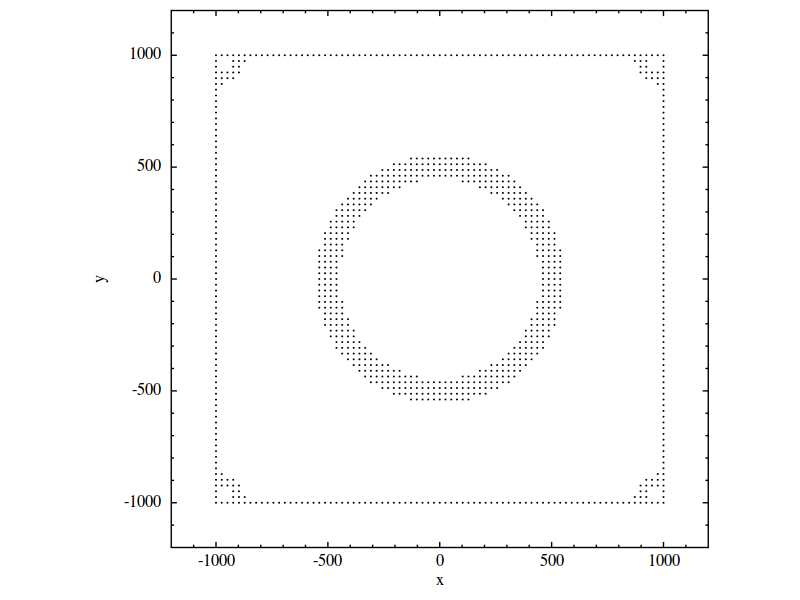} \\
\includegraphics[scale=0.3]{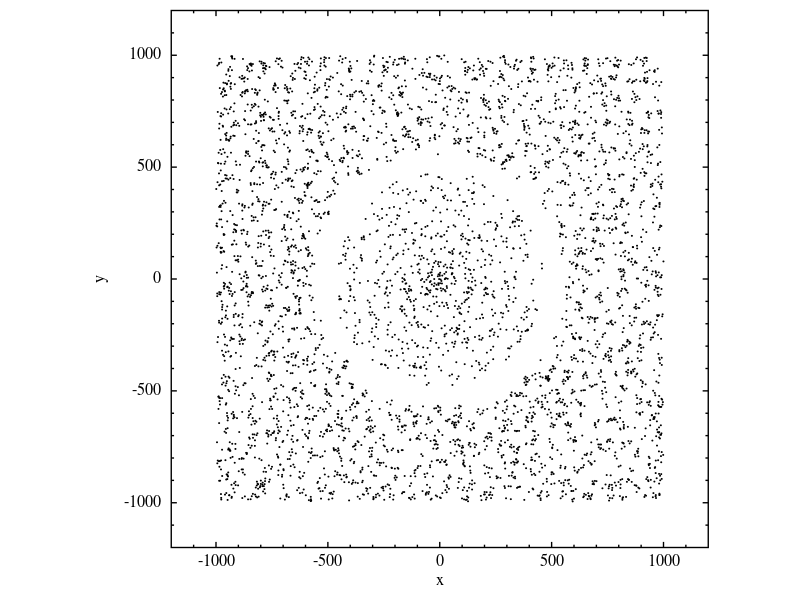} &
\includegraphics[scale=0.3]{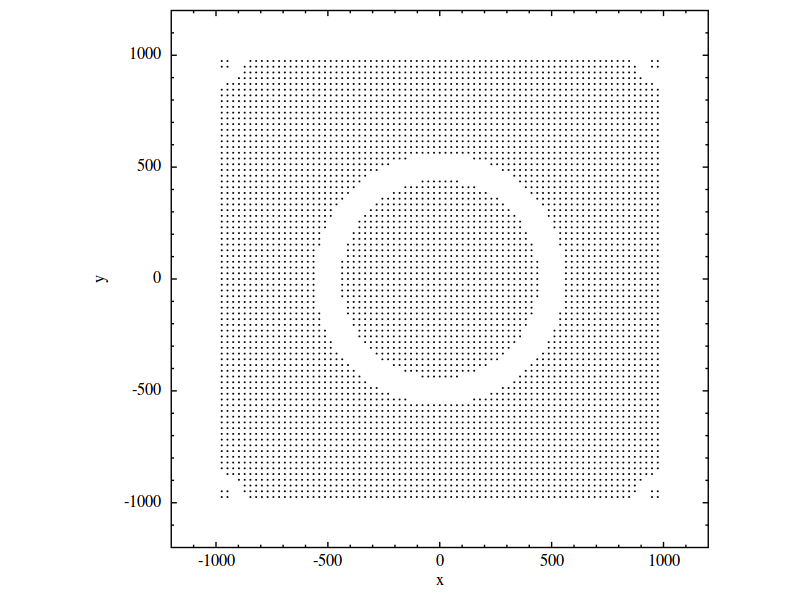} \\
\end{array}$
\caption{The distribution of particles by class in Case 1 (left) and Case 2 (right), for a threshold parameter of $\lambda_T=10^{-6}$.  The top plots show particles classified as sheets, and the bottom row particles classified as voids.  \label{fig:sheet_plots}}
\end{center}
\end{figure*}

\noindent Figure \ref{fig:sheet_plots} shows the distribution of sheet particles (top row) and void particles (bottom row) for the random and lattice configurations (left and right respectively).  In both cases, the tensor classification identifies the surface of the expanding shell, which has a thickness of order the local smoothing length, as is evidence by the void classifications. Also, both cases show that particles inside the sphere, more than a smoothing length from the velocity change at $R=500$, are classified as voids.  However, in Case 1, substantial particle noise ensures that the velocity flow inside and outside the sphere is not precisely uniform, resulting in a number of erroneous sheet classifications outside of the shell, due to small fluctuations in the velocity gradient.  In Case 2, no such problem arises, and the classification is essentially perfect, even capturing the sheet nature of particles at the boundary of the cube.

\section{Applications}\label{sec:applications}

\subsection{Spiral Structure in Discs \label{sec:disc}}

\noindent Very young protostellar discs are thought to undergo a brief self-gravitating phase.  During this epoch, the Toomre parameter \citep{Toomre_1964}:

\begin{equation}
Q = \frac{c_s \Omega}{\pi G \Sigma} \lessapprox 1.5,
\end{equation}

where $c_s$, $\Omega$ and $\Sigma$ are the local sound speed, angular velocity and surface density respectively.  As a result, non-axisymmetric instabilities can grow in the disc \citep{Durisen_review}.  These manifest as spiral density waves, with the number of spirals and their properties being largely determined by the nature of how angular momentum is transported \citep{Forgan2011}.  

In extreme circumstances, cool, extended self-gravitating discs can fragment into bound objects (see e.g. \citealt{Rice_et_al_05}).  While the precise conditions for fragmentation are the subject of debate \citep{Meru2011,Lodato2011,Forgan2011a,Rice2014, Young2015}, once these conditions are met fragmentation tends to begin in the spiral arms themselves, where the local density is significantly higher than the mean at that radius.  Even when fragmentation does not occur, these spiral structures are the agents of angular momentum redistribution (and consequently mass accretion) in the disc. It has also been shown that spiral density waves are capable of accumulating dust via gas drag, and enhancing grain growth, a result with obvious implications for subsequent planet formation via core accretion \citep{Rice2004,Clarke2009,Dipierro2015}.  Therefore, an important task for simulators of self-gravitating discs is to understand the growth and evolution of these spiral arms.

\begin{figure*}
\begin{center}$\begin{array}{cc}
\includegraphics[scale=0.3]{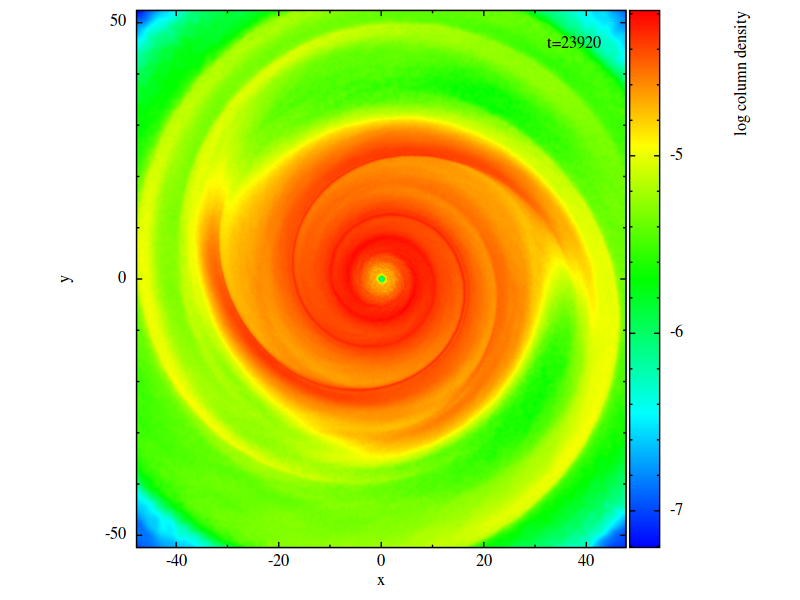} &
\includegraphics[scale=0.3]{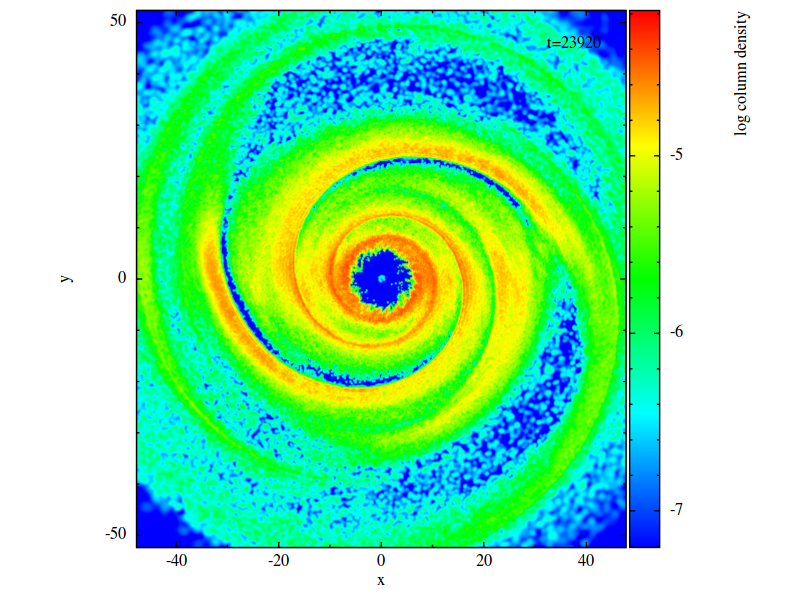} \\
\includegraphics[scale=0.3]{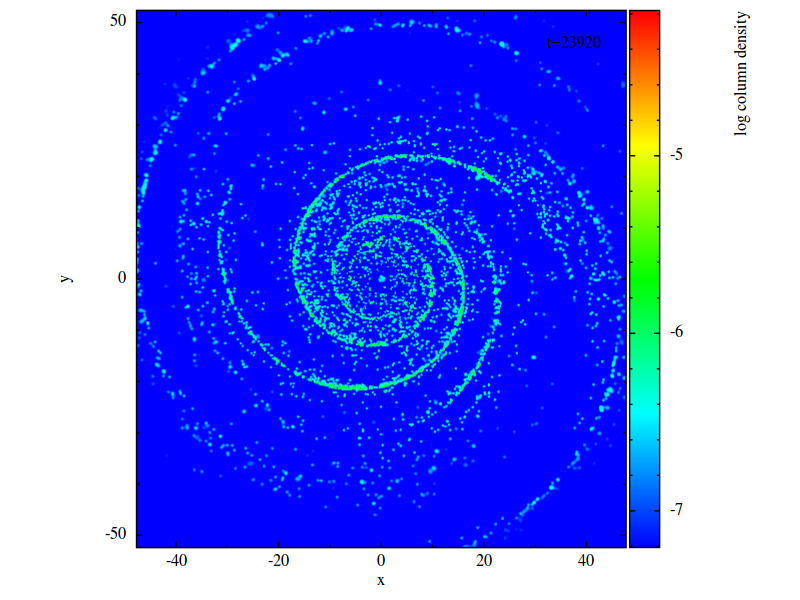} &
\includegraphics[scale=0.3]{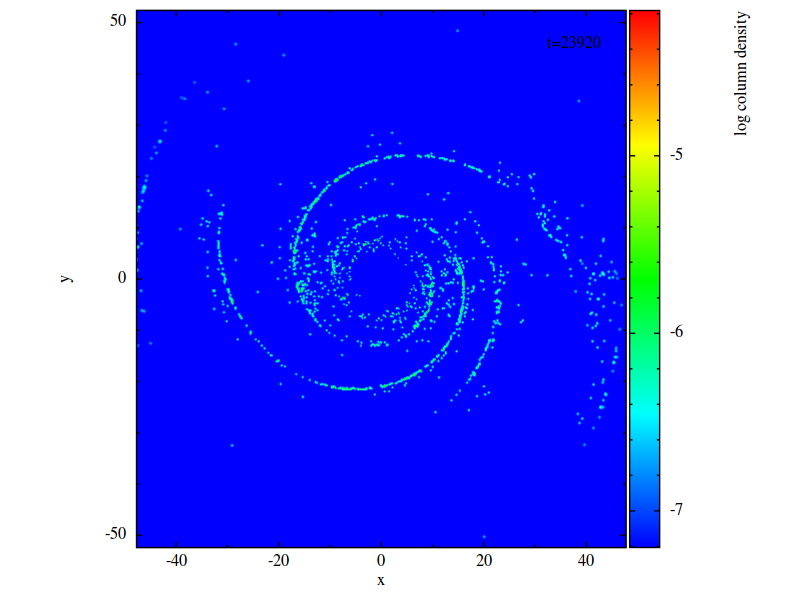} \\
\end{array}$
\caption{The classification of disc spiral arms. Top left: the complete SPH simulation of a self-gravitating protostellar disc.  Top right: Particles classified as clusters using the tidal tensor.  Bottom Left: Particles classified as clusters using the velocity shear tensor.  Bottom Right: Particles classified as clusters using both tensors. \label{fig:discplots}}
\end{center}
\end{figure*}

Our tensor classification methods can be used to great effect here. The top left panel of Figure \ref{fig:discplots} shows an SPH simulation of a self-gravitating protostellar disc, which has settled into a quasi-steady state.  The star mass is 1 $\msol$, and the disc mass is 0.25 $\msol$.  Radiative transfer is approximated using the hybrid formalism of \citet{intro_hybrid}.  The disc is isolated, i.e. it does not accrete from an envelope, and is sufficiently compact to avoid fragmentation.  Generation of spiral structures, which produces weak shock heating, is balanced by radiative cooling, resulting in a self-regulated state where $Q$ is close to the instability value.  

Classification of cluster particles using the tidal tensor (top right panel of Figure \ref{fig:discplots}) identifies the arms quite well, due to their relatively strong density perturbation above the mean.  This tells us which particles are inside the arm at this moment in time.  

It might seem odd that classification of filaments is not used to detect the arms.  While filament particles do tend to broadly trace the arms, the long range potential of the arms ensures that many of the disc's particles are classified as filaments even when at relatively large distances from any wavefront, rather than as sheets, which might be more intuitive given the disc's geometry.  However, when classifying using the velocity shear tensor, as we do below, it is indeed the case that most particles are identified as sheets.  

The velocity shear tensor probes the local velocity gradient, and so is more effective at resolving the wavefront, as can be observed in the bottom left panel of Figure \ref{fig:discplots}.  This measure is more useful for measuring the arms' shape, and with multiple snapshots can be used to estimate the pattern speed of the wave.  

Finally, the results of both classifications can be combined, where we now demand particles to share the same cluster classification for both tidal and velocity shear tensors (bottom right panel).  This measures the wavefront of the ``strong'' waves, removing particles from the outer regions and reducing the thickness of the wavefront itself.  This approach can run the risk of removing too many particles from the simulation, leaving too few to conduct a sensible analysis.  However, if one is only interested in waves that are particularly strong, this ``correlated-tensor'' approach can be quite useful.

\subsection{Structures in Molecular Clouds \label{sec:SN}}

Star formation in galaxies is mediated by the giant molecular clouds (GMCs), which subsequently undergo localised collapse where the gas is sufficiently cool and dense.  The evolution of structure in these molecular clouds is governed by the interplay of gravitational attraction, pressure (both hydrodynamic and magnetic), radiative and kinetic feedback from ongoing star formation and stellar death, and larger scale external forcing from galactic scale structures such as spiral arms (see \citealt{Dobbs2014a} for a review).

The competition between the above forces results in complex, turbulent structures.  In particular, the identification of active star formation in filaments (e.g. \citealt{Andre2010, Hacar2011}) chimes well with numerical simulations of star formation, which show an abundance of filamentary structures (e.g. \citealt{Klessen2000,Federrath2010,Krumholz2011,Bonnell2013, Moeckel2015}).  The precise character of the filamentary structure derived from simulations will depend on what physical processes are active.

We show an example here of the effect of supernova feedback on the structure of a molecular cloud.  The initial conditions are based on from Run I of \citet{Dale2012}.  While the aforementioned investigated the effects of photoionising feedback, we use a control run, which was not subjected to such effects.  A $10^4 \msol$ molecular cloud, of radius $10$ pc, was initially represented by $10^6$ particles.  The cloud is initially globally unbound, seeded with supersonic turbulence with $v_{rms} = 2.1 \mathrm{km \,s^{-1}}$.  The gas undergoes cooling according to the cooling function of \citet{VazquezSemadeni2007}.

As the goal was to simulate supernova feedback, which requires finer resolution, each SPH particle was split into nine daughter particles, giving a mass resolution of $1.1\times10^{-3}\msol$ per particle.  The accretion radius of each sink particle was subsequently reduced to $10^{-3} pc$.

The most massive sink in the simulation - with mass $95.6 \msol$ - was selected as supernova progenitor.  One quarter of the sink's mass was assumed to become ejecta in the explosion, and hence the sink's mass was reduced by this amount.   To simulate the ejecta, a sphere of 21,507 particles was evolved in a separate, isolated simulation to a (more) relaxed state by including only pressure forces and light damping to reduce gradients in density and pressure. The SN radius was 0.01 pc, with a hole at the centre matching the sink's accretion radius at $10^{-3}$ pc. The ejecta has a total energy of $10^{51}$ ergs.  Half of this energy is thermal, while the rest is kinetic, with a velocity that is purely radial.  

\begin{figure*}
\begin{center}$\begin{array}{cc}
\includegraphics[scale=0.3]{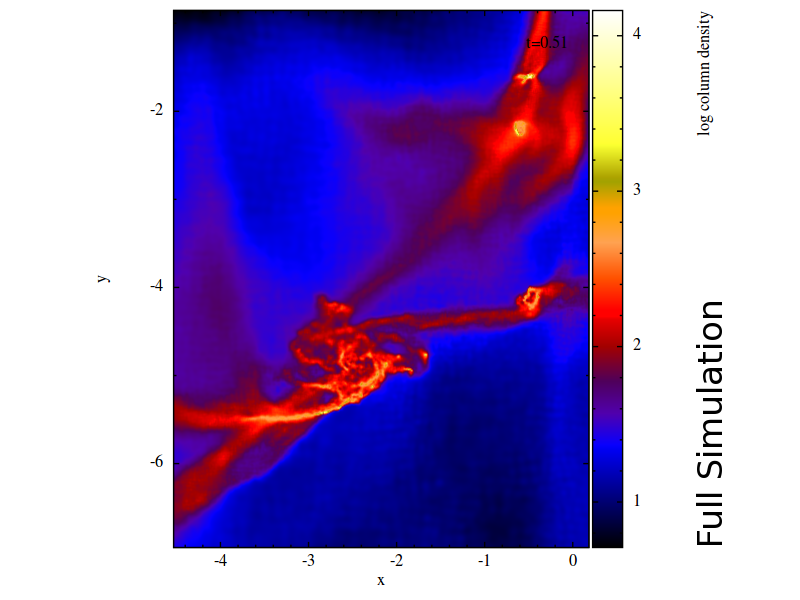} &
\includegraphics[scale=0.3]{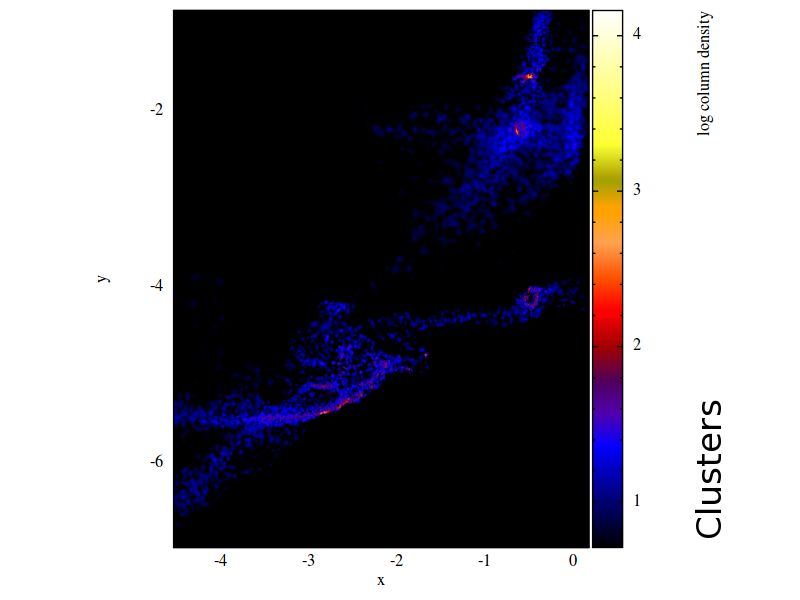} \\
\includegraphics[scale=0.3]{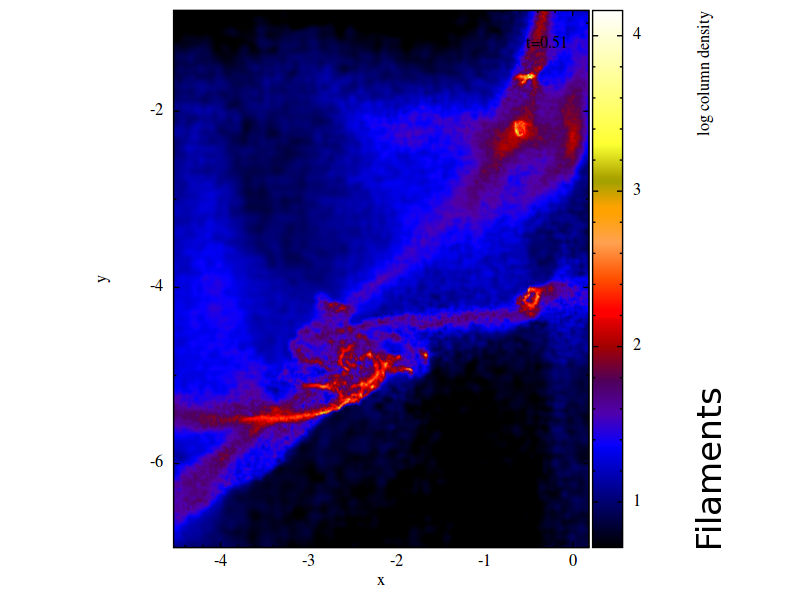} &
\includegraphics[scale=0.3]{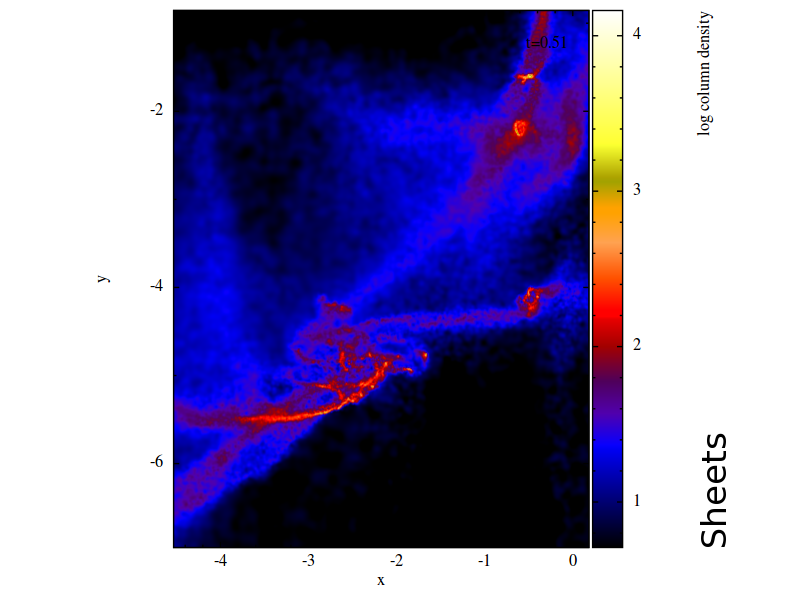} \\
\end{array}$
\caption{Particle classifications in the SPH simulation (using the tidal tensor) without a supernova explosion. Top left: the complete SPH simulation.  Top right: particles classified as clusters.  Bottom left: particles classified as filaments.  Bottom right: particles classified as sheets. \label{fig:noSN}}
\end{center}
\end{figure*}

We therefore have two simulations to investigate - one containg a supernova, and one that does not.  For both simulations, we only consider a subset of the particles in the vicinity of the supernova's location.  Firstly, we consider the simulation without a supernova, and analyse its structure.  Figure \ref{fig:noSN} shows the simulation (top left), and its particles classified as clusters (top right), filaments (bottom left) and sheets (bottom right) using the tidal tensor.  As is expected, cluster particles trace the densest regions of gas, in the process of collapsing to form stars.  The filament classifications do indeed trace filamentary structures.  The sheet particles possess a very similar distribution to the filaments, suggesting that the filaments are embedded within sheets.  Comparison of the local eigenvectors shows that for the velocity shear tensor, the angle between filament and sheet normals has a mode of approximately $\pi/2$, suggesting that filaments are typically embedded within the plane of the sheet.  However, measurements of this angle using the tidal tensor shows a similar angle distribution, but with a mode of approximately $\pi/3$.

Table \ref{tab:SN} shows the relative change in classifications as a result of the supernova being added to the simulation.  For the tidal tensor, we can see that the supernova tends to produce more cluster and filament classifications, with a decrease in the number of sheet and void counts.  Before the supernova, approximately 9\% of the simulation is classified as cluster, with around 48\% identified as filament.  A small fraction of the simulation (around 4\%) are classified as voids, so the relatively strong change in void fraction should be considered carefully.

The velocity shear tensor shows quite different behaviour, with an increase in particles identified as sheets, which makes intuitive sense given that the supernova produces an expanding shell of gas akin to the tests conducted in section \ref{sec:velsheartest}.  The increase in sheet identifications (as well as a small increase in cluster identifications) suggests that the supernova encourages the compression of gas, and therefore provides weak assistance in instigating star formation.  The decrease in filament classifications may be due to the sweeping up of particles on the periphery of filaments into the blast wave of the supernova, which is consistent with both the increase in sheet classifications, and also the increase in void classifications.

\begin{table}
\centering
  \caption{The proportions of various particle classifications using the tidal tensor (T) and the velocity shear tensor (V) before and after adding a supernova explosion\label{tab:SN}}
  \begin{tabular}{c || cc || cc}
  \hline
  \hline
   Class & $f_{noSN}$ (T)  &  $f_{SN}$ (T) & $f_{noSN}$ (V)  &  $f_{SN}$ (V) \\
   \hline
Cluster   & 9.4 \%   & 10.4 \%  & 1.9\%  & 2.0\%     \\
Filament & 48.6\% & 50.4\%   & 55.0\% & 51.3\%  \\
Sheet      & 37.8\%  & 35.2\%   & 42.3\% & 45.6 \% \\
Void        & 4.2\%    & 4.0 \%   & 0.7 \%  & 1.1\%     \\
 \hline
  \hline
\end{tabular}
\end{table}

\begin{figure*}
\begin{center}$\begin{array}{cc}
\includegraphics[scale=0.4]{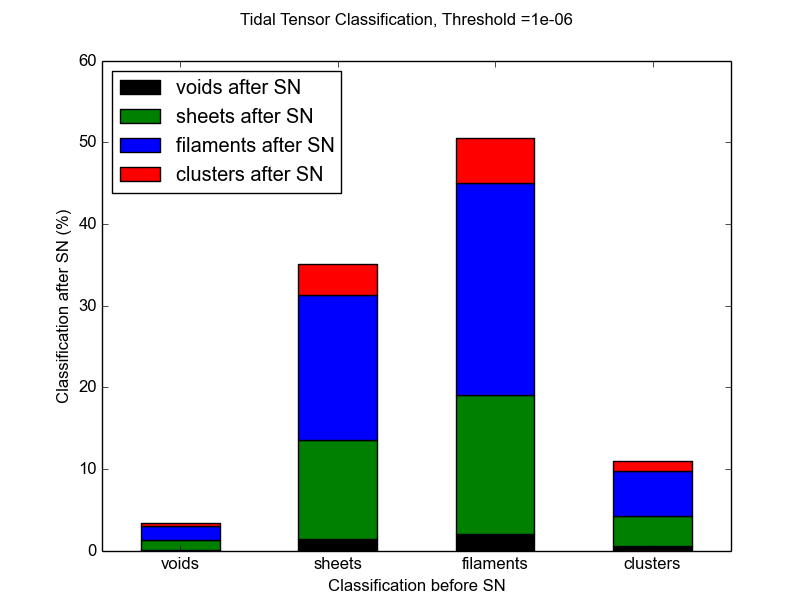} &
\includegraphics[scale=0.4]{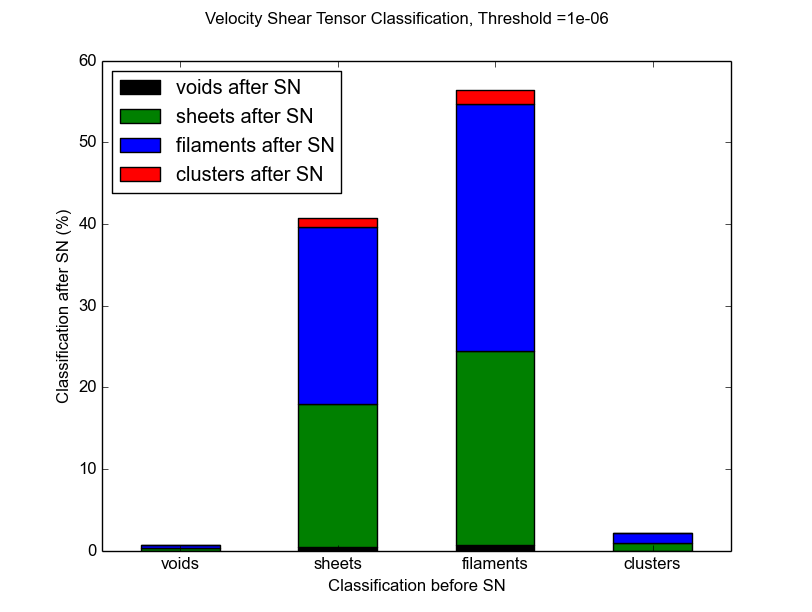} \\
\end{array}$
\caption{The breakdown of particle classifications before and after the supernova.  The total height of each vertical bar indicates the population of particles as they were initially classified before the supernova.  Each bar is then broken down into sections indicating how the population was re-classified post-supernova.  The left plot shows the data for the tidal tensor, and the right plot the data for the velocity shear tensor. \label{fig:SNbar}}
\end{center}
\end{figure*}

Figure \ref{fig:SNbar} gives a breakdown of how particles of different types are subsequently re-classified.  The total height of each vertical bar gives the fraction of each population before the supernova occurs, and the height of each section of the bar shows what fraction of each component ends up as a given structure type.  For example, the total height of the sheets bar for the left panel of Figure \ref{fig:SNbar} is 37.8\%, corresponding to the total given in Table \ref{tab:SN}.  Only a third of these particles remain as sheets post-supernova, with the others changing type to the other three possible classes.

This plot allows us to ask, for example, whether adding a supernova sculpts clusters from sheets or filaments.  In the case of the tidal tensor (left plot) we can see that most of the particles classified as clusters post-supernova (the top sections of each vertical bar) were originally filaments, although a significant fraction were also originally sheets.  Many particles were filaments before the supernova and sheets after, and vice versa. Almost all of the particles classified as voids after the supernova were not voids to begin with!

The velocity shear tensor (right plot) shows some similar trends, although the number of void particles is significantly reduced.  The number of sheet particles increases as a result of the supernova, although the filament population remains the dominant component of the simulation, as is confirmed by Table \ref{tab:SN}.

\begin{figure*}
\begin{center}$\begin{array}{cc}
\includegraphics[scale=0.3]{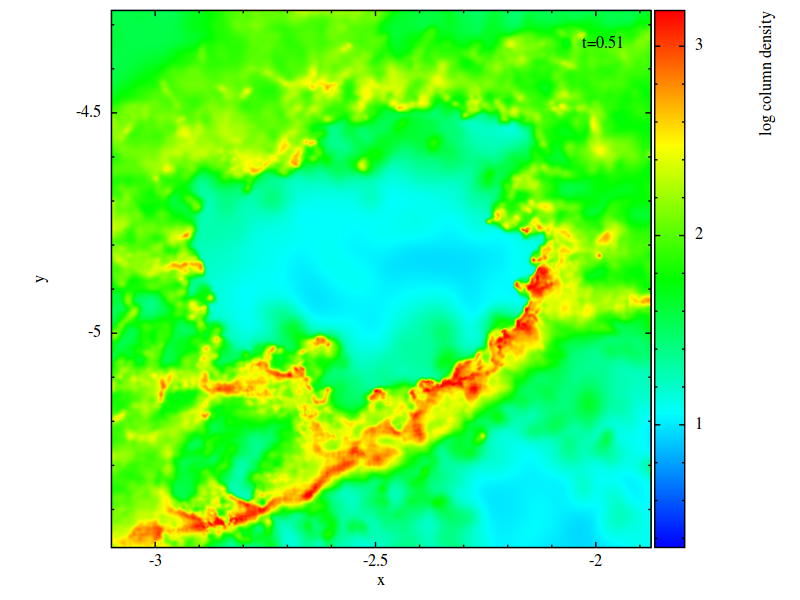} &
\includegraphics[scale=0.3]{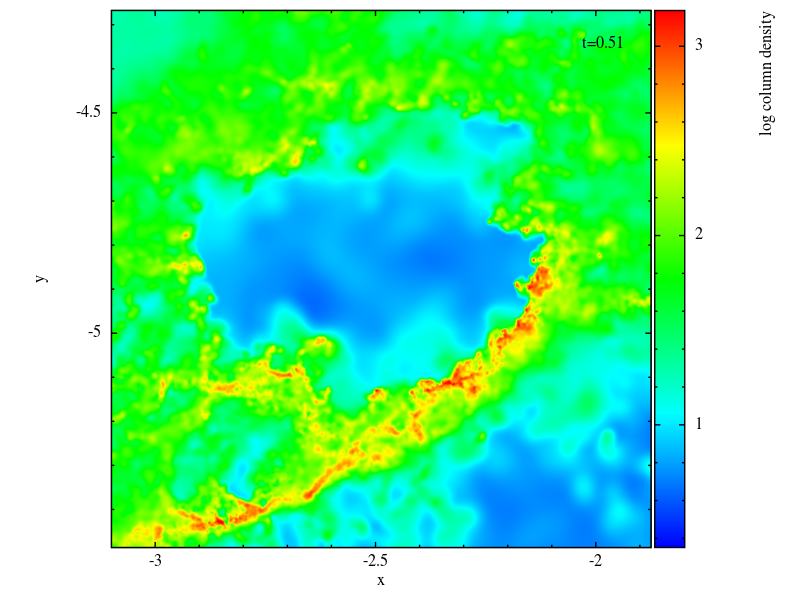} \\
\includegraphics[scale=0.3]{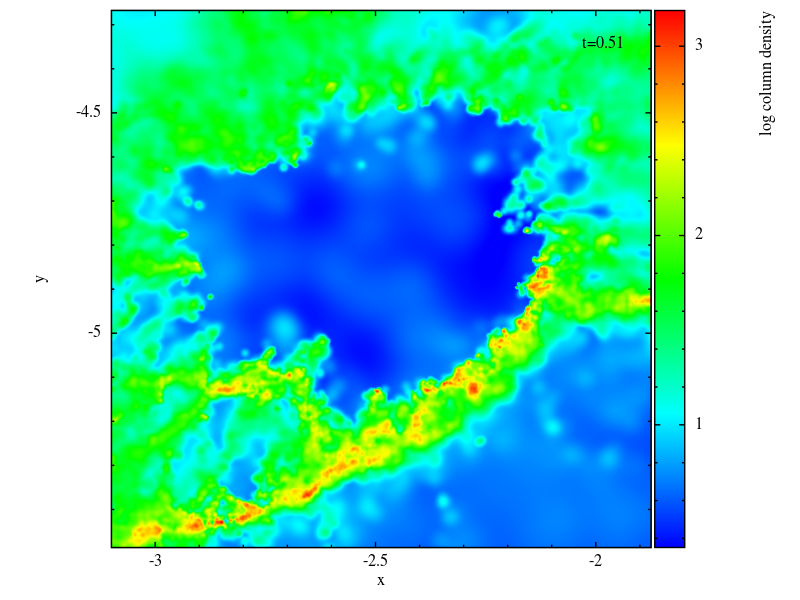} &
\includegraphics[scale=0.3]{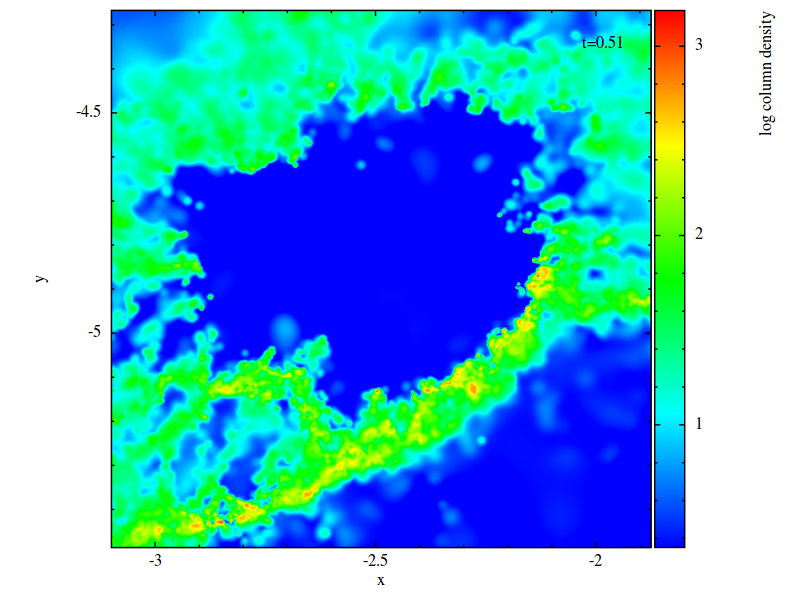} \\
\end{array}$
\caption{The classification of filaments during a supernova explosion.  Top left: A subset of the complete SPH simulation.  Top right: particles classed as filaments using the tidal tensor. Bottom left: particles classified as filaments using the velocity shear tensor. Bottom right, particles classified as filaments simultaneously with both tensors. \label{fig:SNfilament}}
\end{center}
\end{figure*}

Figure \ref{fig:SNfilament} shows the effect of the supernova on the surrounding material.  The top left plot shows all particles in the vicinity of the blast.  We now test our ``correlated-tensor'' approach from the previous example, where we now identify filaments using the tidal tensor (top right plot), the velocity shear tensor (bottom left plot), and both tensors simultaneously (bottom right).  This approach has allowed us to identify the blast-front of the supernova.  The tidal tensor identifies particles at the inner edge of the cavity driven by the supernova, as these have interacted with the wave for a sufficient amount of time to be swept into dynamically quasi-stable structures.  The velocity shear tensor probes the particles that have just begun their interaction with the blast-front.  They are yet to be swept into self-gravitating structures, although the presence of a relatively strong filament in the vicinity of the blast does limit our ability to interpret this.

The correlated tensor approach again gives further insight.  As the two tensors probe different regions of the blast, looking for simultaneous classifications is in effect looking for regions where the blast is still sweeping matter into high velocity flows, while at the same time collapsing into semi-bound structures.  It traces the densest regions of the filaments seen in the top left plot of Figure \ref{fig:SNfilament}, and may also be tracing sites of future star formation triggered by the supernova - further simulation is required to confirm this hypothesis.

\section{Discussion} \label{sec:discussion}

\subsection{Limits of the Method}

\noindent We have seen that tensor classification can be exceptionally powerful when applied in the correct circumstances.  Equally, we have seen that like all classification schemes, it can be hampered by numerical effects.  

The greatest issue appears to be large amounts of particle noise.  As previously stated, SPH is typically formulated to reduce noise as the simulation progresses \citep{Monaghan_05,Price2012}, so provided that the number of particles is sufficiently large in the regions of interest, it seems that this problem can be surpassed, particularly if one considers the simulation after sufficient dynamical times have elapsed.  The velocity shear tensor appears to be less sensitive to particle noise, so we recommend its deployment in noisy circumstances. 

Perhaps more pressing in practice is the issue of overlapping structures, an issue common to all classification schemes.  As we saw in section \ref{sec:SN}, identifying the causal agents of structure when it impinges on pre-existing structures is challenging.  In these cases, classifying multiple snapshots of a simulation is the best antidote.  For the sake of brevity, we have not investigated the time evolution of structures.  As SPH is a purely Lagrangian method, it is appropriate to track a particle's classification as a function of time, which would allow us to investigate (for example) what structures a particle passes through before it participates in sink formation, or accretion.  The supernova particles in section \ref{sec:SN} could have their classifications traced to characterise the state of the blast front with time.  As particles pass through spiral structures like those in section \ref{sec:disc}, the physical properties of the particle can be studied.  If the disc fragments, the particles that constitute the fragment will contain a classification history that will provide insights on fragment formation and evolution. 

The classification method as it stands operates on a particle-by-particle basis, and as such only gives information on  structure at the location of each particle, and does not give information about the connectivity of structures.  For example, we can identify regions of the simulation that are filamentary structures, but we do not currently identify whether the region is a single filament or composed of multiple interlinked filaments, such as filament bundles detected in observations and simulations of star formation \citep{Hacar2013, Smith2015}.  Further algorithms are required to decompose a filamentary region into its constituent sub-filaments.  CLUMPFIND and its cousins may be suitable for this purpose. 

As with all SPH simulations, structures cannot be produced below the local smoothing length, and clearly sub-$h$ structures cannot be detected using this technique.  The dimensionless nature of the tensors does allow the method to probe close to the resolution limit of the simulation, but no further.  Care should therefore be taken when classifying regions of a simulation that are not well-populated by SPH particles, as these particles will typically have large smoothing lengths, and hence be limited in their ability to produce structure.

\subsection{Prospects for Future Work}

The concept of tensor classification is not limited to the two tensors we have investigated in this paper.  Investigation of other properties of the medium allow us to construct other tensors, whose eigenvalues and eigenvectors will provide insights into the system's structure\footnote{Unfortunately, the inertia tensor cannot be employed in this fashion, as its eigenvalues are always positive.}.

In the case of magnetohydrodynamics (MHD), the Maxwell stress tensor may be of use:

\begin{equation}
\sigma_{ij} = \frac{1}{\mu_0} \left(B_i B_j - \frac{1}{2}B^2 \delta_{ij} \right),
\end{equation}

\noindent Where we have quoted its ``E-free'' form, suitable for ideal MHD, and $B_i$ and $\mu_0$ are the magnetic field components and permeability of free space respectively.  This tensor has the units of pressure, and like other stress tensors, its eigenvectors are the normals of planes along which the shear component of the field is zero (the ``principal stresses'').  Again, the sign of the eigenvalues indicates tension (positive) or compression (negative).  Tensor classification of the magnetic field therefore indicates the dimension of manifold the magnetic field is attempting to create.  While magnetic flux is ``frozen in'' to the medium in ideal MHD, comparison of the magnetic field topology to the gravitational potential's topology via the tidal tensor may deliver new insight into how these fundamental forces compete and collaborate to produce bound and unbound objects.

Finally, it seems clear that this formalism is extendable to special and general-relativistic SPH (e.g. \citealt{Siegler2000,Monaghan2001, Rosswog2010}).  If we consider the Newtonian tidal tensor $T_{ij}$, differential geometry indicates that the mean local curvature in the potential surface 

\begin{equation}
H= 0.5 \,\mathrm{Tr}(T_{ij}) = \frac{1}{2} \sum_i \lambda_i,
\end{equation}

\noindent and the Gaussian curvature

\begin{equation}
K = 0.5 \prod_i \lambda_i.
\end{equation}

\noindent This intrinsic relationship between the tidal tensor, the curvature of the potential surface, and the local matter density distribution, is quite analogous to the relationship between mass-energy density and the curvature of space-time as captured by Einstein's field equations \citep{Misner1973,Masi2007}.  In general relativity, the analogous quantity is the Riemann tensor $R^i_{jkl}$, where in the weak field limit

\begin{equation}
R^i_{00j} = T_{ij}.
\end{equation}

\noindent The rank 2 Ricci tensor $R_{ij} = R^k_{ikj}$ might therefore be an encouraging tensor for classification.  $R_{ij}$ represents the deviation of ``spherical'' volumes from their Euclidean equivalent due to local space-time curvature, and as such the eigenvalues and eigenvectors of this tensor will provide useful information.  Precisely how this may be done, whether it will bear any resemblance to the positive-eigenvalue classification system previously described, and how applicable it may be in SPH simulations with a fixed metric, is beyond the scope of this paper, but worth further investigation.

\section{Conclusions}\label{sec:conclusions}

\noindent In this paper, we have described how tensor classification methods, initially used for $N$-Body simulations, can be used on smoothed particle hydrodynamics (SPH) density fields to identify structures.  Classification of the eigenvalues of either the tidal tensor or the velocity shear tensor at the point of an SPH particle provide local information on how matter is collapsing or flowing respectively, in particular what stable manifold is being produced.  

The sign of the tensor eigenvalues, in particular the number of positive eigenvalues ($E$) indicates the dimension of this stable manifold, and hence we can classify the topology of the structures in the simulation accordingly.  In the case of 1D (2D) manifolds, the eigenvector corresponding to the positive (negative) eigenvalue gives directional information.  For 1D manifolds (sheets), this gives the normal vector of the sheet plane, and for 2D manifolds (filaments), this gives the flow direction vector of the filament.

Identifying structures in this fashion allows us to a) identify regions of interest in very large simulations, and perform robust statistical analyses that can then be compared with observations, and b) compare and contrast the effects of different physical processes to drive structure formation by classifying using multiple tensors.

SPH density fields are generally smooth and continuous, allowing tensor classification to be carried out with less parametrisation than in the $N$-Body case, and without requiring tesselation of the density field, as used by some algorithms \citep{Sousbie2011}.  However, due to floating point error and other numeric issues, we must retain one free parameter, the eigenvalue threshold $\lambda_T>0$.  Eigenvalues which exceed $\lambda_T$ are classified as ``positive'', and hence $E$ is the number of eigenvalues greater than $\lambda_T$.  

We have shown through simple tests that the tidal tensor and velocity shear tensor can reproduce analytically derived test examples, provided that the particle disorder is relatively low, which is likely to be the case for simulations that have been evolved on scales that are long compared to the dynamical time.  Classifications using the velocity shear appear to be less sensitive to this disorder.

There are many possible use cases for such a technique - we have outlined only two.  Spiral arms in self-gravitating discs can be easily identified using either tensor, with each tensor revealing different aspects of the spiral density wave (the wavefront, the total number of particles entrained in the wave, etc.).  Filamentary structure in molecular clouds can be easily discerned using the tidal tensor, and the effects of feedback can be quantified at a structural level.

We believe that these techniques will prove to be extremely useful in SPH simulations at a variety of scales, and that these classification techniques are not limited to the two tensors discussed in this work.  Indeed, we advocate further study of other tensors as tools for identifying other types of structure present in SPH data, as this is likely to yield new and fine-grained insight, even as simulations continue to grow in scope and complexity.  

\section*{Acknowledgments}

DF, IB and WL gratefully acknowledge support from the "ECOGAL" ERC advanced grant.  KR gratefully acknowledges support from STFC grant ST/M001229/1.  This work relied on the compute resources of the St Andrews MHD Cluster.  Surface density plots were created using \texttt{SPLASH} \citep{SPLASH}.  The authors thank the anonymous referee for their insightful comments, which strengthened this manuscript.

\bibliographystyle{mn2e} 
\bibliography{sph_tensor}

\label{lastpage}

\end{document}